\documentstyle[preprint,eqsecnum,aps]{revtex}
\begin{document}
\draft
\title{Nucleus-Electron Model for States Changing from \\
a Liquid Metal to a Plasma and the Saha Equation
}
\author{J. Chihara and Y. Ueshima}
\address{Advanced Photon Research Center, Japan Atomic Energy Research Institute\\
Tokai, Ibaraki 319-1195, Japan}
\author{S. Kiyokawa}
\address{Department of Physics, 
Nara Women's University, \\
Kita-Uoya Nishimachi, Nara 630-8506, Japan}
\date{\today}
\renewcommand{\baselinestretch}{2.0}
\maketitle

\begin{abstract}
We extend the quantal hypernetted-chain (QHNC) method, 
which has been proved to yield accurate results for liquid metals, 
to treat a {\it partially ionized} plasma. 
In a plasma, the electrons change from a quantum to 
a classical fluid gradually with increasing temperature; 
the QHNC method applied to the electron gas is in fact able to 
provide the electron-electron correlation at arbitrary temperature.
As an illustrating example of this approach, we investigate 
how liquid rubidium becomes a plasma  
by increasing the temperature from 0 to 30 eV 
at a fixed normal ion-density $1.03\!\times\!10^{22}/{\rm cm^3}$.
The electron-ion radial distribution function (RDF) in liquid Rb 
has distinct inner-core and outer-core parts. Even at a temperature of 
1~eV, this clear distinction remains as a characteristic of 
a liquid metal. At a temperature of 3 eV, this distinction disappears, 
and rubidium becomes a plasma with the ionization 1.21. 
The temperature variations of bound levels in each ion and 
the average ionization are calculated in Rb plasmas at the same time.
Using the density-functional theory, we also derive the Saha equation applicable even to a high-density 
plasma at low temperatures. 
The QHNC method provides a procedure to solve this Saha equation 
with ease by using a recursive formula; 
the charge population of differently ionized species are 
obtained in Rb plasmas at several temperatures.
In this way, it is shown that, with the atomic number as the 
only input, the QHNC method produces the average ionization, 
the electron-ion and ion-ion RDF's, and the charge population which are 
consistent with the atomic structure of each ion for 
a partially ionized plasma.
\end{abstract}
\vspace{1cm}

\section{INTRODUCTION}\label{intro}
In order to calculate thermodynamic functions, transport coefficients 
and optical properties in a partially ionized plasma, 
it is a fundamental problem to determine 
the average ionization $Z_{\rm I}$, the equilibrium correlations 
among ions and electrons, the atomic structure 
(bound-levels in the ions) and the charge population (ionization balance) 
of differently ionized species, in a self-consistent way 
with each other in these quantities. 
However, at the present stage there is no theory which can 
produce these quantities 
in a {\it partially ionized} plasma in a unified manner.
It is the purpose of the present work to show that the quantal 
hypernetted-chain (QHNC) equation\cite{QHNC} developed 
for a liquid metal can be extended to calculate these quantities of 
a partially ionized plasma in a unified manner. 

Up to the present, in the calculation of thermodynamic functions 
or optical properties in a partially ionized plasma, 
the ion-sphere (IS) model\cite{tfISr,ISLib,tfISs,INFERNO,Qeos,IS} is 
used as the standard method.
Although there are many kinds of variations in the IS model, 
the essential point of this model is that the ion-ion correlation 
in a plasma is approximated by 
the step function $\theta(r\!-\!a)$ with the Wigner-Seitz radius $a$; 
an atom is considered to be either confined within the ion-sphere 
or immersed in the infinite jellium.
It should be remarked that this model is only applicable for 
high-density and low-temperature systems, that is, limited to a 
narrow range of densities and temperatures for the plasma state.

On the basis of the density functional (DF) theory, 
Dharma-wardana and Perrot\cite{DP} derived a set of integral equations for 
static correlation functions in a strongly coupled plasma; it was shown\cite{jcDP} 
that this method breaks down when it treats a plasma with a significant number 
of bound electrons as an ion due to its improper treatment of the electron-ion 
correlation. To judge whether a theory treating a partially 
ionized plasma is proper or not, 
we have the criterion as to what extent it can reproduce 
the observed structure factors of liquid metals 
for which many reliable experimental data exist. 
This is a liquid metal can be taken as a special type of partially 
ionized plasma. 
In this context, we have proposed a set of integral equations 
for radial distribution functions (RDF) in a liquid metal 
as a nucleus-electron mixture\cite{QHNC} 
on the basis of the DF theory in the quantal hypernetted-chain (QHNC) 
approximation. At this point, it should be kept in mind that the QHNC theory 
is limited to treating only a ``simple" liquid metal, where the bound states 
are clearly distinguished from the continuum state. 
Already, we have applied the QHNC method to several simple 
liquid metals\cite{QHLi,QHNa,QHK,QHAl,fpMD}, 
and  obtained their structure factors 
in excellent agreement with experiments. 
In these calculations, we have demonstrated that 
the QHNC method can determine the ``outer structure" 
(the ion-ion and electron-ion RDF's and the ionic charge $Z_{\rm I}$) 
in a consistent way with the ``inner structure" (the atomic structure 
of the ions) using the atomic number $Z_{\rm A}$ of the system 
as the only input data. Therefore, the QHNC method is suited for 
treating a plasma, where the ion-ion and electron-ion interactions  
may vary over a wide range in conjunction with the internal structure 
of each ion according to change of state condition. 
In a similar spirit, Perrot\cite{NPA} has proposed 
the neutral-pseudoatom (NPA) 
method based on the DF theory to calculate 
the effective ion-ion potential for a partially ionized plasma. 
Gonz\'alez {\it et al.} have successfully applied 
this theory to alkali liquids\cite{gonzalk} and 
alkaline-earth liquids\cite{gonzearth}. The NPA method can 
be derived from the QHNC theory with additional use of 
the IS approximation\cite{jcDP} in the determination of 
the pseudopotential to construct an effective ion-ion interaction. 
This model is called in our approach the jellium-vacancy model\cite{QHNa} 
and is used to obtain an initial guess for the effective 
ion-ion interaction in the iteration to solve the QHNC equation 
for a liquid metal. At this point, it should be recognized that 
the NPA method is not appropriate for treating a high-temperature 
plasma with the weak ion-ion correlation, since this method is 
based on the IS model; this fact will be discussed in the present work.

From successful applications of the QHNC method to liquid 
metals as mentioned before, we can expect this method 
constitutes an adequate representation of 
a partially ionized plasma.
In the present work, we extend the QHNC method applicable 
to a plasma state, where the electrons change from a quantum 
fluid to a classical fluid gradually with increasing temperature; 
this is in contrast with a liquid metal where the electrons 
can be assumed to be perfectly in the Fermi degenerate state 
at zero temperature because of the density being high. 
When the electrons in a plasma begin to behave as classical particles, 
it is difficult to calculate the free-electron density 
distribution under the external potential by solving 
the wave equation. Therefore, Furukawa\cite{furuk} used 
the Thomas-Fermi (TF) approximation to evaluate 
the electron-ion RDF in the QHNC equation for a plasma. 
Also, Xu and Hansen\cite{xu} applied the TF version of 
the QHNC method to a hydrogen plasma with a gradient correction 
to the TF kinetic energy of the electrons.
In the full use of the wave equation to calculate the free-electron 
density distribution, we demonstrate that the QHNC method can treat 
a partially ionized plasma by taking liquid rubidium 
as an illustrating example; this exhibits what changes are found 
when a liquid-metal turns to a plasma state with increasing 
temperature at a fixed ion-density. 
In a liquid metal, there is clear distinction between 
the inner-core and outer-core structures in the electron-ion RDF, 
which allows one to construct an electron-ion pseudopotential 
in a liquid metal. When a liquid metal changes into a plasma state, 
this distinction disappears, and 
we can not set up a peudopotential in the same manner 
as is performed in the usual liquid-metal theory.

In principle, the DF theory generates the exact density distributions of 
electrons and ions in a plasma. Note that 
it can yield only the average-ion structure in a partially ionized plasma. 
In a real plasma, there are many differently ionized ions around the average ion.
The fundamental theory of ionization in a plasma is provided by the Saha equation.
However, the usual Saha equation can be applied only to a low density 
equilibrium plasma, where the interactions among particles are negligible. 
Although there are many modifications of the Saha equation for applicability to dense plasmas by introducing the continuum 
lowering, any modified theory can not treat a dense plasma at low temperatures. 
In the present work, on the basis of the DF theory we derive the Saha equation, 
which provides the charge population of differently ionized ions 
in a dense plasma in the region from low to high temperatures; 
this charge population yields an average ionization consistent 
with an average ion in the plasma determined the DF theory. 
The bound levels and the chemical potential contained 
in the Saha equation are supplied by the QHNC equation 
for the ion-ion and electron-ion RDF's in the plasma. 
In this way, the QHNC method is shown to generate 
the electron-ion and ion-ion RDF's, 
the average ionization $Z_{\rm I}$, and 
the charge population of differently ionized species 
to be consistent with the atomic structure (bound levels) 
of each ion in a unified fashion, as will be shown by the example of a 
rubidium plasma.

The paper is organized as follows. In \S.~\ref{form}, 
we present a summary of the QHNC method 
along with the one-component QHNC equation   
for an electron gas in the jellium model, 
making extensions to treat a plasma. 
As an illustrating example, the application of this formulation 
to a Rb plasma is shown in \S~\ref{num}, where the numerical technique 
to solve the QHNC equation is explained.
In \S~\ref{saha}, we set up the Saha equation 
on the basis of the DF theory, and the charge populations 
are calculated for Rb plasmas at a fixed liquid-metal 
density for several temperatures. 
The last section is devoted to discussion, where the limitation of 
the IS model is also examined, and prospects of applications based on 
the QHNC method are mentioned, such as calculations of the atomic structure and 
transport coefficients in a plasma, and the determination of 
the effective interactions to be used for the molecular-dynamics simulation of a 
plasma as a classical electron-ion mixture.

\section{LIQUID METAL AND PLASMA AS NUCLEUS-ELECTRON MIXTURE}\label{form}

We can set up the following three models to treat liquid metals and plasmas.
In the simplest model, a liquid metal can be treated as a neutral liquid, 
when its interatomic potential is given.
However, this interatomic potential must be introduced from the outside of the model; 
this is impossible when treating a plasma. 
In the second model, a liquid metal is considered
as a binary mixture of ions and electrons; in this model, 
the ionic charge $Z_{\rm I}$ and the electron-ion interaction are unknown 
except for a perfectly ionized plasma\cite{hyd,DP}, 
even if the ion-ion interaction is taken as a pure Coulombic. 
In the case of a plasma, it is rather a fundamental problem to 
determine the ionization $Z_{\rm I}$.
Most fundamental model is to consider 
a liquid metal as composed of nuclei and electrons. 
In this model, all input data are known beforehand 
if provided the atomic number $Z_{\rm A}$ of a liquid metal; 
this first-principles approach enable us to treat a plasma 
in a wide range of temperatures and densities. 
Therefore, let us think of a plasma 
as a nucleus-electron mixture\cite{QHNC} consisting of 
$N_{\rm I}$ nuclei and $Z_{\rm A}N_{\rm I}$ electrons.
Here, we single out one nucleus and 
fix it at the origin; then, the fixed nucleus at the origin in 
the mixture causes the external potentials for electrons and ions, 
and induces an inhomogeneous system. 
This inhomogeneous system can be equivalently translated into a simpler system:
a fixed nucleus with the atomic number $Z_{\rm A}$ is surrounded 
by electrons and ions, of which structure $\rho_{\rm b}(r)$ is 
undetermined at first and should be determined self-consistently at final.
In this simplified model, the fixed nucleus at the origin is 
surrounded by $(N_{\rm I}\!-\!1)$ ions 
interacting via a potential $v_{\rm II}(r)$ and 
by $Z_{\rm I}(N_{\rm I}\!-\!1)\!+\!Z_{\rm A}$ electrons. 
Under this circumstance, the DF theory provides effective external potentials 
$v_{i{\rm N}}^{\rm eff}(r)$ ($i\!\!=\!\!e$ or I), 
which yield the exact electron- and ion-density distributions 
around the fixed nucleus on the basis of 
the following reference system\cite{QDCF}: a mixture composed 
of $(N_{\rm I}\!-\!1)$ noninteracting ions and $Z_{\rm I}(N_{\rm 
I}\!-\!1)\!+\!Z_{\rm A}$ noninteracting electrons. 
Each ion in the reference system is assumed to have 
$Z_{\rm B}$ bound-electrons with a distribution $\rho_{\rm b}(r)$.
The electron-density distribution around the central nucleus is 
determined by solving the wave equation for noninteracting electrons 
in the form:
\begin{equation}\label{eq:df-dns}
n_{\rm e}(r|{\rm N})=n_{\rm e}^0(r|v_{\rm eN}^{\rm eff})=
n_{\rm e}^{\rm b}(r|{\rm N})+n_{\rm e}^{\rm f}(r|{\rm N})\,, \label{e:nen}
\end{equation}
under the effective external potential,
\begin{equation}\label{eq:df-pot}
v_{e{\rm N}}^{\rm eff}(r)=-Z_{\rm A}e^2/r+
{\delta {\cal F}_{\rm int}\over\delta
 n_e(r|{\rm N})}-\mu_e^{\rm int}\,, \label{e:gvef1}
\end{equation}
which is constructed by the DF theory in terms of 
the interaction part of the intrinsic free-energy ${\cal F}_{\rm int}$ 
defined from the reference system and the interaction part 
of the chemical potential $\mu_e^{\rm int}$.
Here, $n_{\rm e}^0(r|U)$ 
is determined by solving the wave equation for an electron 
under the external potential $U(r)$ in the form:
\begin{equation}\label{eq:wave}
n_e^0(r|U)=\Sigma_i f(\epsilon_i)|\psi_i(r)|^2\,,
\end{equation}
where $\psi_i(r)$ is the wave function of the state $\epsilon_i$ (bound or free) 
and $f(\epsilon)$, the Fermi distribution function.  
Now, the density distribution $n_{\rm e}^0(r|v_{\rm eN}^{\rm eff})$ 
can be divided into two parts: the bound- and free-electron 
density distributions, $n_{\rm e}^{\rm b}$  and  $n_{\rm e}^{\rm f}$, 
respectively, according to the states $\epsilon_i$ being negative or positive.

The bound electrons $n_{\rm e}^{\rm b}(r)\!\equiv\!n_{\rm e}^{\rm b}(r|{\rm N})$ 
around the nucleus at the origin 
should be taken to constitute the ion at the origin. Therefore, 
unknown ion structure $\rho_{\rm b}(r)$ in the premise is determined 
by the condition that the central ion with $n_{\rm e}^{\rm b}(r)$ 
formed at the origin must be the same to any surrounding ion 
with the assumed structure $\rho_{\rm b}(r)$ around it, that is, 
\begin{equation}\label{eq:boundsc}
\rho_{\rm b}(r)=n_{\rm e}^{\rm b}(r|{\rm N})\,.
\end{equation}
This leads to a self-consistent condition to determine 
the ion structure $\rho_{\rm b}(r)$; the central ion structure 
$n_{\rm e}^{\rm b}(r)$ is iteratively used as the input $\rho_{\rm b}(r)$ 
giving the new reference system, which determines the next effective potential 
(\ref{eq:df-pot}) to evaluate a new $n_{\rm e}^{\rm b}(r)$.
From this relation, the bound-electron number $Z_{\rm B}$ of 
surrounding ions can be evaluated from the bound-electron density 
distribution $n_{\rm e}^{\rm b}(r|{\rm N})$ 
by $Z_{\rm B}\!=\!\int_0^{\infty}n_e^{\rm b}(r)d{\bf r}$, that is, 
\begin{equation}\label{eq:shzb}
Z_{\rm B}\equiv \sum_{i=1}^M {g_i\over \exp[\beta 
(\epsilon_i-\mu_{\rm e}^0)]+1}\;, 
\end{equation}
for the ion with $M$ bound states with the degeneracy $g_i$, 
and the ionic charge is obtained by $Z_{\rm I}\!\equiv\! Z_{\rm A}\!-\!Z_{\rm B}$.
In addition, the chemical potential $\mu_{\rm e}^0$ involved in 
Eq.~(\ref{eq:shzb}) is determined by noting the fact:
\begin{equation}
 \lim_{r\rightarrow\infty}n_{\rm e}^{\rm f}(r|{\rm I})=\int {2\over \exp[\beta 
(p^2/2m-\mu_{\rm e}^0)]+1} \frac{d{\bf p}}{(2\pi\hbar)^3}=
n_0^{\rm e}=Z_{\rm I}n_0^{\rm I}\,,\label{e:ioniz}
\end{equation}
which can be rewritten in the form: 
\begin{equation}\label{eq:chem}
Z_{\rm A}=\sum_{i=1}^M {g_i\over \exp[\beta 
(\epsilon_i-\mu_{\rm e}^0)]+1} 
+\frac1{n_0^{\rm I}}\int {2\over 
\exp[\beta (p^2/2m-\mu_{\rm e}^0)]+1} 
\frac{d{\bf p}}{(2\pi\hbar)^3}\,.
\end{equation}
Here, $n_0^i$ denotes the number density of electrons or ions ($i=$e or I). 
Also, the bare ion-electron interaction is obtained from 
Eq.~(\ref{eq:df-pot}) with use of some approximations\cite{NEmodel}:
\begin{equation}\label{eq:vei}
v_{\rm eI}(r)  \equiv  -\frac{Z_{\rm A}}{r} +\int v_{\rm ee}^{\rm c}(|{\bf 
r}-{\bf 
r'}|)\,n_{\rm e}^{\rm b}(r')d{\bf r'} +\mu_{\rm XC}(n_{\rm e}^{\rm b}(r)+n_0^{\rm e}) 
-\mu_{\rm XC}(n_0^{\rm e}), \label{e:vei}  
\end{equation}
where $\mu_{\mbox{\tiny XC}}(n_0^e)$ is the
exchange-correlation potential in the local-density approximation.

With use of the average ionization $Z_{\rm I}$ and the bare 
ion-electron interaction $v_{\rm eI}(r)$, 
a plasma can be now modeled as a mixture of electrons and
ions interacting through pair potentials $v_{ij}(r)$ [$i,j=\mbox{e
or I\,}$].  Applying the DF theory to this
electron-ion mixture model, the ion-ion and electron-ion RDF's 
$g_{i\rm I}(r)$ are exactly expressed in
terms of direct correlation functions (DCF) $C_{ij}(r)$ and bridge
functions $B_{i\rm I}(r)$ as follows\,\cite{QHNC}:
\begin{eqnarray}
g_{\rm eI}(r) & = & n_e^{0f}(r|U_{\rm e}^{\rm eff})/n_0^e\;, \label{eq:gii}\\
g_{\rm II}(r) & = & \exp[-\beta U_{\rm I}^{\rm eff}(r)]\;\label{eq:gei}\,,
\end{eqnarray}
with
\begin{eqnarray}
&&U_{i\rm I}^{\rm eff}(r)  \equiv v_{i\rm I}(r)-{\it\Gamma}_{i\rm I}(r)/\beta
-B_{i\rm I}(r)/\beta\;, \label{eq:Ueff}\\
&&{\it\Gamma}_{i\rm I}(r)\equiv \sum_l
\int C_{il}(|{\bf r}-{\bf r}'|)n_0^l[g_{l\rm I}(r')-1]d{\bf r}'\;.\label{eq:gamiI}
\end{eqnarray}
Here, $n_e^{0f}(r|U_{\rm eI}^{\rm eff})$ indicates the free-electron part of 
the density distribution.
We can see a similarity of these expressions to those of a classical
binary mixture.
The difference is found only in that the electron-ion RDF is determined 
by a wave equation in stead of the Boltzmann factor $\exp[-\beta U_{\rm e}^{\rm eff}(r)]$.

Moreover, it is shown from these expressions that the electron-ion mixture 
can be described as a one-component fluid interacting only via 
pairwise interaction $v_{\rm eff}(r)$, if the bridge function $B_{\rm II}(r)$ 
is taken to be the one-component bridge function. 
Namely, the integral equations (\ref{eq:gii})--(\ref{eq:gamiI})
for $g_{i\rm I}(r)$ can be transformed into
 a set of integral equations for the one-component model 
 of plasmas. One is an usual integral equation 
 for the DCF $C(r)$ of a one-component fluid:
\begin{equation}\label{eq:QHNCii}
C(r)=\exp[-\beta v_{\rm eff}(r)
  +{\it\gamma}(r)+B_{\rm II}(r)]-1-{\it\gamma}(r)
\end{equation}
with an interaction $v_{\rm eff}(r)$, and the other is an equation 
for $v_{\rm eff}(r)$, that is expressed in the form of an integral
equation for the electron-ion DCF $C_{\rm eI}(r)$:
\begin{equation}\label{eq:QHNCei}
\hat B C_{\rm eI}(r)
 =n_{\rm e}^{0f}(r|v_{\rm eI}
     \!-\!{\it\Gamma}_{\rm eI}/\beta\!-\!B_{\rm eI}/\beta)/n_0^{\rm e}
     -1-\hat B {\it\Gamma}_{\rm eI}(r)\;,
\end{equation}
since the effective interionic interaction $v_{\rm eff}(r)$ is given by
\begin{equation}\label{eq:veff}
\beta v_{\rm eff}(Q)\equiv\beta v_{\rm II}(Q)
 -{|C_{\rm eI}(Q)|^2 n_0^{\rm e} \chi_Q^0\over
 1-n_0^{\rm e} C_{\rm ee}(Q) \chi_Q^0 }\;.
\end{equation}
Here, ${\it\gamma}(r)\equiv \int C(|{\bf r}-{\bf r}'|)n_0^{\rm I}[g_{\rm II}
(r')-1]d{\bf r}'$, and
$\hat B$ denotes an operator defined by
\begin{equation}
{\cal F}_Q[\hat B^\alpha f(r)]
  \equiv (\chi_Q^0)^\alpha{\cal F}_Q[f(r)]
  =(\chi_Q^0)^\alpha\int\exp[i{\bf Q}\cdot{\bf r}]f(r)d{\bf r}\;,
\end{equation}
for an arbitrary real number $\alpha$, and represents a
quantum-effect of the electrons through the density response
function $\chi_Q^0$ of the noninteracting electron gas. 
We can obtain a set of closed integral equations (referred to 
as the QHNC equation) from Eqs.~(\ref{eq:QHNCii})--(\ref{eq:veff}) by
introducing the following approximations\,\cite{QHNC}. 
(1) $B_{\rm eI}\simeq 0$ (the HNC approximation). 
(2) The bridge function $B_{\rm II}$ of the ion-electron mixture is
approximated by that of one-component hard-sphere fluid 
(modified HNC approximation\cite{VMHNC}).
(3) An approximate $v_{\rm eI}(r)$ is taken to be Eq.~(\ref{eq:vei}), 
where we adopt the Gunnarsson-Lundqvist formula\cite{GL} for 
the exchange-correlation potential $\mu_{\mbox{\tiny XC}}(n_0^e)$. 
(4) $v_{\rm II}(r)$ is taken as a pure Coulombic $Z_{\rm I}^2e^2/r$.  
(5) For liquid metals where the electrons can be assumed 
to be at zero temperature, the electron-electron DCF is approximated by 
$C_{\rm ee}(Q)\simeq -\beta v_{\rm ee}(Q)[1-G^{\rm jell}(Q)]$
in terms of the the local-field correction (LFC) 
$G^{\rm jell}(Q)$ of the jellium model; we have used the LFC proposed 
by Geldart and Vosko\cite{GV} in our many applications to liquid metals.

\begin{figure}
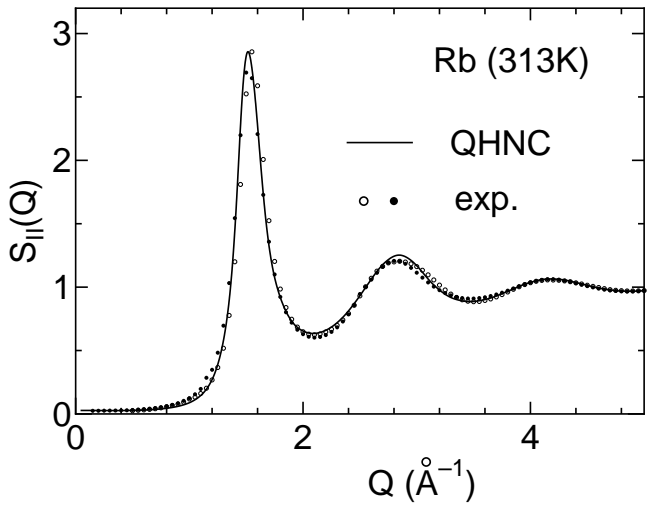

\caption{Ion-ion structure factor $S_{\rm II}(Q)$ for liquid Rb 
at a temperature of 313 K; the QHNC method yields a structure factor (full curve) 
in excellent agreement with experiments (open\protect\cite{ExpRbCH} and full circles\protect\cite{ExpRbW}).}
\label{fig:SQ313}
\end{figure}

Under these approximations, a set of integral equations can be solved 
to determine the electron-ion and ion-ion correlations in a liquid metal 
together with the ionization and electron bound states.
Figure \ref{fig:SQ313} is an applied example of this procedure to liquid metals: 
the structure factors of liquid rubidium at temperature 313 K 
and density $1.03\!\times\!10^{22}/{\rm cm}^3$.
The full curve is the QHNC result, which exhibits an excellent agreement 
with experiments denoted by open circles 
(neutron-scattering\cite{ExpRbCH}) and full circles (X-ray\cite{ExpRbW}).
From this example, we can expect the QHNC method to yield good results for 
partially ionized plasmas.\par
However, when we apply the QHNC method to a plasma, there occur the following
problems;
the LFC involved in the electron-electron DCF must be evaluated 
at arbitrary temperature in dealing with a plasma. 
Although the LFC at the absolute zero temperature has been calculated 
by many investigators and applied to many kinds of liquid metals, 
there is no standard way to calculate the LFC at finite temperature.
For this purpose, we adopt the one-component QHNC equation\cite{QHNCone} for 
an electron gas in the uniform positive background to obtain 
the electron-electron DCF at arbitrary temperature, 
which is written for an integral equation for the electron-density 
distribution $n_{e}(r|e)$ around the fixed electron in an electron gas, 
\begin{equation}\label{eq:QHNCee}
n_{e}(r|e)=n_e^0(r|U_{\rm eff})=\Sigma_i f(\epsilon_i)|\psi_i(r)|^2
\end{equation}
with
\begin{equation}\label{eq:QHNCU}
\beta U_{\rm eff}(r)\equiv\beta v_{ee}(r)\!-\!\int C_{ee}(|{\bf r}\!
-\!{\bf r}'|)[n_{e}(r'|e)\!-\!n_0^e]d{\bf r}'\,. 
\end{equation}
Here, the DCF for a one-component system is defined by
\begin{equation}\label{eq:QHNCdcf}
n_0^eC_{ee}(Q)\equiv 1/\chi_Q^0-1/\chi_Q^{ee}=-\beta v_{ee}(Q)[1-G(Q)]\,. 
\end{equation}
The Fourier transform of the density distribution yields the following
bootstrap relation to determine the DCF $C_{ee}(Q)$ with combined use 
of Eqs.~(\ref{eq:QHNCee}) and (\ref{eq:QHNCU}):
\begin{equation}\label{eq:QHNCboot}
{\cal F}_{\! Q}[n_e(r|e)-n_0^e]=\chi_Q^{\rm ee}/\chi_Q^0-1
=1/[1\!-\!n_0^e C_{ee}(Q)\chi_Q^0]-1\,, 
\end{equation}
which is derived from a certain ansatz\cite{QHNCone}. At this point, 
it should be noted that the QHNC equation~(\ref{eq:QHNCee}) 
reduces to the well known HNC 
equation for a classical electron gas
in the high temperature limit, because of the relations: 
$\chi_Q^0\!\!=\!\!1$, $\chi_Q^{\rm ee}\!\!=\!\!S_{\rm ee}(Q)$ 
and $n_e^0(r|U_{\rm eff})\!=\!n_0^e\exp[-\beta U_{\rm eff}(r)]$ in the classical limit.
As a result, the approximation (5) can be replaced by  
the one-component QHNC equation (\ref{eq:QHNCee}) 
for the electron gas;  with use of other approximations 
(1)--(4), this replacement makes Eqs~(\ref{eq:QHNCii})--(\ref{eq:veff}) 
a closed set of equations for a plasma including a liquid metal as a special case.
The electron-ion and the ion-ion RDF's of liquid Rb at 313 K are plotted in Fig.~\ref{fig:iieiRDF} 
along with the effective ion-ion interactions 
calculated using both the QHNC and Geldart-Vosko $G(Q)$; the resulting 
two effective potentials differ with each other, but yield almost the same 
ion-ion RDF's as shown by the full curve and open circles. 
In Fig.~\ref{fig:iieiRDF} the electron-ion RDF obtained from the QHNC method 
has an inner-core structure which is caused by the orthogonality 
of the free-electron to the bound-electron wave functions.
On the other hand, it should be noticed that the usual liquid-metal theory 
based on the Ashcroft pseudopotential\cite{ashcroft} yields 
an electron-ion RDF, which has no inner-core structure 
(shown by full circles). The Ashcroft pseudopotential is constructed 
by neglect of this inner-core structure; 
this cut-off of the inner-core structure brings about a simple treatment 
of liquid metals in the standard liquid-metal theory.

\section{NUMERICAL CALCULATION APPLIED TO RUBIDIUM PLASMA}\label{num}
Already, we have calculated the electronic and ionic structures of liquid rubidium 
in a wide range of temperatures and densities: 
compressed states\cite{JCGK} and expanded states\cite{SCCS97}.
To show the applicability of the QHNC method to a plasma state, therefore here 
we take, as an example, liquid rubidium changing its temperature 
from 313 K to 3.5$\times 10^5$ K (30 eV) at fixed ion-density 
$r_{\rm s}^{\rm I}\!=\!5.388$. 
Here, $r_{\rm s}^{\rm I}$ denotes the Wigner-Seitz radius 
$a\!\equiv\! r_{\rm s}^{\rm I}a_{\rm B}$ 
in units of the Bohr radius $a_{\rm B}$.

In the application of the QHNC equation to a plasma, we must calculate 
the electron-density distribution $n_e^0(r|U_{\rm eff})$ 
under the external potential $U_{\rm eff}(r)$ to get $g_{\rm eI}(r)$ and 
$n_{\rm e}(r|{\rm e})$ by solving the wave equation generally. 
However, it is difficult to determine 
this density distribution from the wave equation at high temperatures, 
where it becomes nearly 
the classical Boltzmann factor:
 $n_e^0(r|U_{\rm eff})=\Sigma_i f(\epsilon_i)|\psi_i(r)|^2 
\Rightarrow n_0^e\exp[-\beta U_{\rm eff}]$.
In the calculation of the free-electron density distribution, 
$n_e^{0f}(r|U_{\rm eff})=2\int f(\epsilon_{\bf p})|
\psi_{\bf p}(r)|^2 {d{\bf p}}/{(2\pi\hbar)^3}$, 
at finite temperature, the electron kinetic energy $\epsilon_{\bf p}$ is not limited 
within the Fermi energy $E_{\rm F}$, as is the case at zero temperature. 
In addition, since the electron-density distribution 
$n_e^0(r|U_{\rm eff})$ begins to 
approach the Boltzmann factor from the large distance 
as the temperature increases, we must calculate 
the wave functions with large angular momentum $l$ to correctly obtain the classical 
electron-density distribution in the large distance at high temperature.
We can circumvent this difficulty by using the Thomas-Fermi (TF) 
approximation to the electron-density distribution for $r>r_{\rm c}$; 
the distance $r_{\rm c}$ can be chosen according to the temperature 
so that $n_e^0(r|U_{\rm eff})$ calculated from the wave equation becomes 
almost equal to the TF result for $r>r_{\rm c}$.
This situation can be seen in the following calculation.

\begin{figure}
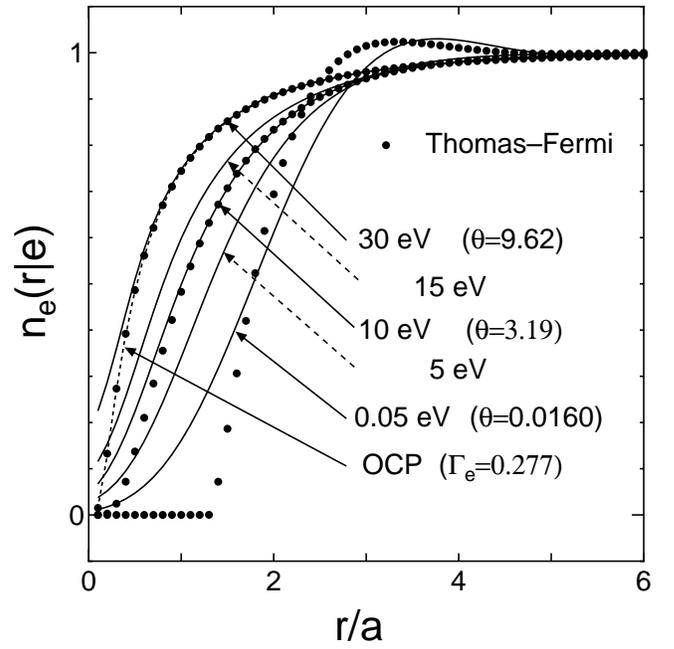

\caption{Electron-electron correlations in an electron gas 
at the density of $r_{\rm s}\!=\!4$
for temperatures ranging from 0.05 to 30 eV 
with $\theta\!\equiv\! k_{\rm B}T/E_{\rm F}$ indicating the electron degeneracy.
The solid curves are calculated by the one-component QHNC equation, 
while the full circles denote the result from 
the TF approximation for each temperature. 
The classical HNC equation for the OCP corresponding to a 
temperature of 30 eV provides the electron-electron correlation 
plotted by the dashed curve (\lq\lq OCP"), which is indistinguishable from the TF 
result at a temperature of 30 eV.
}
\label{fig:rs4}
\end{figure}

The electron-density distribution around a fixed electron is calculated 
for a partially degenerate electron plasma at the density 
$2.51\!\times\!10^{22}/{\rm cm}^3\,$, that is, $r_{\rm s}\!=\!4$ 
in terms of $r_{\rm s}a_{\rm B}\equiv (3/4 \pi n_0^e)^{1/3}$ defined 
for the electron density $n_0^e$; the results calculated for temperatures 
from 0.05 eV to 30 eV are shown in Fig.~\ref{fig:rs4}. 
The electron degeneracy is denoted by $\theta\equiv k_{\rm B}T/E_{\rm F}$, 
i.e., the temperature over the Fermi energy.
For high degeneracy (0.05 eV), the TF approximation 
(denoted by full circles) gives quite a different 
density distribution from the one  calculated by the wave equation.
When the temperature is increased to 10 eV, 
the TF result becomes almost same to that obtained by the 
wave equation except near the origin.
When the temperature approaches 30 eV, 
the electron-electron correlation reduces to the classical one; 
the classical HNC equation for the one-component plasma (OCP) with 
the electron plasma parameter $\Gamma_{\rm e}=0.277$ (30 eV) provides 
an indistinguishable result from the TF calculation as is shown 
in Fig.~\ref{fig:rs4} by the dashed curve. Here, the electron 
plasma parameter is defined by $\Gamma_e\equiv \beta e^2/r_{\rm s}a_{\rm B}$.
In the calculation of $n_e(r|e)$, we can obtain 
at the same time the electron-electron DCF, 
which determines the LFC $G(Q)$ from the relation 
$\beta C_{\rm ee}(Q)\!=\!-v_{\rm ee}(Q)[1\!-\!G(Q)]$; 
the calculated $G(Q)$ at $r_s\!=\!4$ 
is shown in Fig.~\ref{fig:Gq} for each temperature corresponding 
to Fig.~\ref{fig:rs4}. This figure indicates that 
the LFC at high temperature can be approximated by that calculated from 
the classical HNC for the OCP, as is shown by the case of 30 eV 
($\theta=9.62$), where the LFC of the classical OCP is denoted by 
the dashed curve. 
In this way, by means of the QHNC equation for an electron gas 
we can obtain the electron-electron DCF, which determines 
the plasma properties in terms of the QHNC $G(Q)$.

Using now the QHNC-LFC $G(Q)$ instead of a 
Geldart-Vosko\cite{GV} type $G(Q)$, 
we apply Eqs.~(\ref{eq:QHNCii})-(\ref{eq:QHNCei}) to rubidium 
at the fixed density of the normal liquid metal; the
temperature has been varied from 0 (313 K) to 30 eV in order to investigate 
how a liquid-metal becomes a plasma.
For the purpose to obtain initial data of the 
ion-core structure and the electron-ion correlation 
for the fully self-consistent QHNC method, we take the jellium-vacancy 
model as a first step. 
In this model the following two approximations are introduced 
in the expression for the electron-ion interaction (\ref{eq:Ueff}), that is, 
\begin{equation}\label{eq:Uei}
\beta U_{\rm eI}^{\rm eff}(r) = \beta v_{\rm eI}(r)- 
\sum_l\int C_{el}(|{\bf r}-{\bf r}'|)n_0^l[g_{l\rm I}(r')-1]d{\bf r}'\;.
\end{equation}
(1) The ion-ion RDF involved above is approximated
by the step function: $g_{\rm II}(r)\!=\!\theta(r\!\!-\!\!a)$, and (2) 
the electron-ion DCF, by a pure Coulomb force: 
$C_{\rm eI}(r)\!=\!-\beta v_{\rm eI}^c(r)$. 
Then, the problem to determine the electron-ion RDF becomes identical with 
the problem to determine the electron density distribution around a fixed nucleus 
at the center of the spherical-vacancy in the jellium background. 
This model is essentially same to the INFERNO model\cite{INFERNO,IS} (or the ion-sphere model) introduced by Liberman and also to 
the neutral-pseudoatom model\cite{NPA} proposed by Perrot. 
By solving these integral equations, we can obtain the electron-ion DCF, 
and the ion-ion effective interaction (\ref{eq:veff}) 
with the combined use of the electron-electron DCF determined 
by the one-component QHNC equation~(\ref{eq:QHNCee}). 
As a consequence, we can obtain initial input data for the fully 
self-consistent QHNC equations 
(\ref{eq:QHNCii})--(\ref{eq:veff}). After that, this set of 
integral equations in conjunction with Eqs.~(\ref{eq:QHNCee}) and (\ref{eq:QHNCU}) 
is solved iteratively by varying the bound-electron number 
$Z_{\rm B}\!=\!\int_0^{\infty}\rho_{\rm b}(r)d{\bf r}$ until the self-consistent 
condition (\ref{eq:boundsc}), $n_{\rm e}^{\rm b}(r)\!=\!\rho_{\rm b}(r)$, is fulfilled.

In the iteration to solve the QHNC equation, we must evaluate also the free-electron 
density distribution around the nucleus at arbitrary temperature. 
It has been the standard method\cite{fTF,TFying,tfISr,tfISs,IS,furuk} 
for the calculation of this 
free-electron density distribution to use the TF approximation in such a way: 
\begin{equation}\label{eq:TFdns}
 n_e^{0f}(r|U)\approx 2\int_{{\bf p}^2/2m+U(r)>0}[\exp(-\beta ({\bf p}^2/2m\! +\! 
U(r)\!-\!\mu_e^0))\!+\!1]^{-1}\!\frac{d{\bf p}}{(2\pi\hbar)^3}\equiv n_f^{\rm TF}(r|U)\;.
\end{equation}
In this expression, the integration of ${\bf p}$ is limited within the domain 
${\bf p}^2/2m\!+\!U(r)>0$ to define the free-electron part of the usual 
TF formula, $n^{\rm TF}(r|U)$, 
which involves both the bound- and free-electron density distributions.
In Fig.~\ref{fig:TFg}, the electron-ion RDF's at 
temperatures of $3.5\!\times\!10^4$ K and $3.5\!\times\!10^5$ K
calculated by the QHNC method are shown in comparison with the results 
from the TF approximation (\ref{eq:TFdns}). It should be noticed that 
the TF formula for the free-electron density distribution 
(\ref{eq:TFdns}) is not a good approximation except for large distances even at a 
high temperature of 3.5$\times 10^5$ K ($\theta\!\!=\!\!5.21$), while there the total 
(free and bound) density distribution can be fairly well described by 
the TF formula. Noting this fact, we can evaluate the free-electron 
density distribution function $n_e^{0f}(r|U)$ for large distances 
$r>r_c$ by the value of $n^{\rm TF}(r|U)\!-\!n_e^{0b}(r|U)$ 
using the TF density distribution $n^{\rm TF}(r|U)$ and the 
bound-electron density distribution 
$n_e^{0b}(r|U)$ obtained by the wave equation; in the case 
3.5$\times 10^5$ K, the cut-distance $r_c$ can be chosen to be 
$0.9a$ as shown by the arrow in Fig.~\ref{fig:TFg}. In this way, 
we can determine the the electron-ion RDF i.e., $n_e^{0f}(r|U)$, 
by solving the wave equation only for small angular momentum $l$; 
it is enough to take the maximum angular momentum $l_{\rm max}=15$ 
even for the high temperature region.

The electron-ion RDF's at temperatures of 1 eV and 3 eV are 
plotted in Fig.~\ref{fig:1ev3ev} along with the ion-ion RDF's 
and the effective ion-ion interactions. 
As is shown in Fig.~\ref{fig:iieiRDF}, 
the electron-ion RDF at a temperature of 313 K has 
distinct inner-core and outer-core parts.
Even at a temperature of 1 eV as shown in Fig.~\ref{fig:1ev3ev}, 
this clear distinction remains as characteristic for a liquid metal; 
the ionization is practically unity
and the ion-ion effective interaction is almost the same structure 
as that of liquid metal at the normal condition, 
although the ion-ion correlation becomes weak because of the small plasma 
parameter $\Gamma\!\equiv \!Z_{\rm I}^{5/3}\Gamma_e$. 
Therefore, we can consider that rubidium  remains as a liquid-metal state 
even at $1.16\!\times\!10^4$ K (1 eV).
Figure~\ref{fig:1ev3ev} shows that at a temperature of 3 eV 
the distinction between inner- and outer-core parts near $0.4a$
disappears and that the ionization, now 1.21,  has become significant.
Because of the disappearance of this distinction between the 
the inner- and outer-core structures, it is difficult to construct a 
pseudopotenital in a plasma state. This makes a contrast with a liquid-metal state, 
where a pseudopotential such as the Ashcroft potential can be used to set up 
an effective interaction between ions in a liquid metal.

\begin{figure}
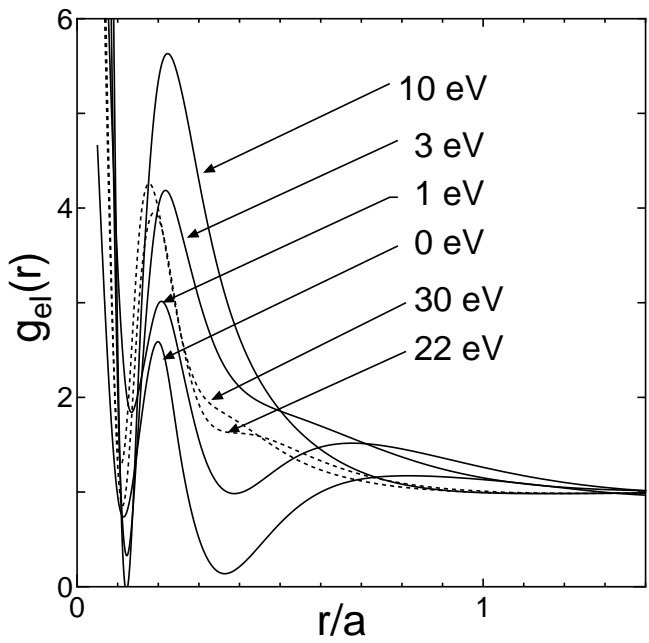

\caption{The temperature dependence of the electron-ion RDF 
for a range from 0 to 30 eV. The inner-core structure near 0.12 reflects 
the variation of the bound-electron wave functions in an ion 
due to the orthogonality between 
the free- and bound-electron wave functions.}
\label{fig:ALLgei}
\end{figure}

\begin{figure}
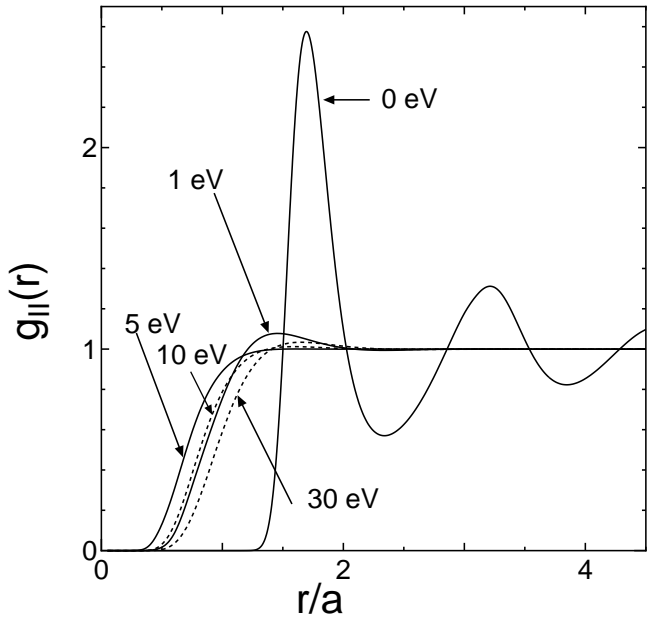

\caption{The temperature dependence of the ion-ion RDF for a range 
from 0 eV (313 K) to 30 eV. The ion-ion RDF becomes small at 1 eV; 
nevertheless rubidium remains as a liquid state with $Z_{\rm I}\!=\!1$.}
\label{fig:ALLgii}
\end{figure}

\begin{figure}
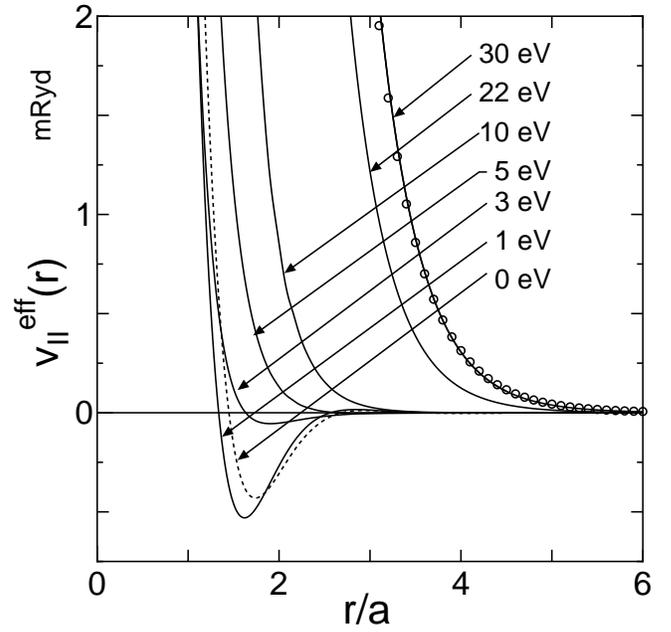

\caption{The temperature dependence of the effective ion-ion interaction 
for a range from 0 eV (313 K) to 30 eV.
The effective potential at a temperature of 30 eV approaches the 
screened Debye potential denoted by the open circles.}
\label{fig:ALLpot}
\end{figure}

The temperature variation of the electron-ion RDF is shown in Fig~\ref{fig:ALLgei} 
for a range from 0 eV to 30 eV. The electron-ion 
correlation becomes stronger for temperatures up to 10 eV, and turns 
to become weaker from 10 eV to 30 eV; the distinction between the inner- and outer-core parts near the point $0.4a$ is gradually disappearing 
with increasing temperature. On the other hand, 
the ion-ion RDF's are shown in Fig.~\ref{fig:ALLgii} 
for increasing temperatures from 0 eV (313 K) to 30 eV. 
Also, the effective ion-ion interactions generating the RDF's 
in Fig.~\ref{fig:ALLgii} are plotted in Fig.~\ref{fig:ALLpot}, 
where the open circles denote the screened Debye potential 
$\exp(-r/D_e)Z_{\rm I}^2e^2/r$ at a temperature of 30 eV 
($\Gamma_e\!\!=\!\!0.277$) with $D_e\equiv (4\pi e^2\beta n_0^e)^{1/2}$.
We can see that the ion-ion effective potential approach the screened Debye potential 
as the electron plasma parameter $\Gamma_e$ becomes small with increasing temperature.
The HNC equation for a one-component fluid with 
the screened Debye potential at a temperature of 30 eV 
provides the RDF $g_{\rm II}^{\rm SD}(r)$ in fair agreement 
with the result from the QHNC method for the electron-nucleus 
mixture; for example, the structure factor at zero 
wavenumber becomes $S_{\rm II}(0)\!=\!0.18$ from the screened Debye potential, which 
should be compared with the QHNC result, $S_{\rm II}(0)\!=\!0.16$.
This fact suggests that in the high temperature region where 
$S_{\rm II}(0)$ becomes large, the ion-sphere model 
(the jellium-vacancy model) can be improved by using 
the approximation $g_{\rm II}(r)\!\approx\! g_{\rm II}^{\rm SD}(r)$ 
instead of the step function.

Figure~\ref{fig:levels} shows the temperature variation of 
the outer-bound levels of an average ion in Rb plasma at the fixed 
ion-density of normal liquid metal. 
The 4s- and 4p-bound levels are plotted there corresponding 
to free atom, 0, 1, 3, 5, 10, 22, and 30 eV, respectively. 
As the temperature increases, the bound levels become deeper 
due to the decrease of bound-electron number, 
which makes the screening effect weak. 
At temperatures of 22 and 30 eV, new bound levels, 5s and 4d, appear.
The occupation number $f(\epsilon_i)$ at the level 
$\epsilon_i$ is written at each level line 
in Fig.~\ref{fig:levels}. 
The ionization variation $Z_{\rm I}$ is shown in the top of 
Fig.~\ref{fig:levels} as the temperature is increased.\\

\section{THE SAHA EQUATION IN THE DENSITY-FUNCTIONAL THEORY}\label{saha}
The DF theory provides the exact electron-density 
distribution $n_e(r|U)$ in 
the nonuniform electron system caused by an external potential $U(r)$; 
however, it is important to notice that this exact density 
distribution $n_e(r|U)$ at finite temperature is 
only an average density distribution. Consider a nucleus 
with the atomic number $Z_{\rm A}$ fixed in an electron gas. 
The effective external potential $U_{\rm eff}(r)$ 
defined by the DF theory gives an exact expression for 
the electron density distribution 
in terms of a density distribution of noninteracting electrons in the form:
\begin{equation}\label{eq:SHdns}
n_{\rm e}(r|{\rm N})=n_{\rm e}^0(r|U_{\rm eff})=
n_{\rm e}^{\rm b}(r|{\rm N})+n_{\rm e}^{\rm f}(r|{\rm N})\;,
\end{equation}
which defines an average ion with the $Z_{\rm B}$ bound-electrons: 
$Z_{\rm B}\!=\!\int n_{\rm e}^{\rm b}(r|{\rm N})d{\bf r}$. 
In the realistic system, the ion should have some integer-number of 
the bound-electrons with fluctuations in time; the number $Z_{\rm B}$ 
given by the DF theory is only an average value of 
this bound-electron number over time. 
The similar situation is found in the ion structure in a nucleus-electron mixture.
Here, we investigate the charge population of differently ionized species in a plasma on the basis of 
the DF theory.

As is discussed in \S.~\ref{form}, the average bound-electron number $Z_{\rm B}$ is 
defined by Eq.~(\ref{eq:shzb}) in the nucleus-electron model based on the DF theory, 
and the chemical potential $\mu_{\rm e}^0$ is determined by Eq.~(\ref{eq:chem}).
This average bound-electron number $Z_{\rm B}$ in an ion in a plasma can be represented by 
\begin{equation}\label{eq:SHZBd}
Z_{\rm B} = \lambda {d \over d\lambda}\ln \Xi_{\rm B}
=\sum_{i=1}^M {g_i \over \exp[\beta(\epsilon_i-\mu_{\rm e}^0))]+1}\,, 
\end{equation}
if we introduce here the grand partition function 
$\Xi_{\rm B}\! \equiv\! \prod_{i=1}^M[1\!+\!\lambda \exp(-\beta \epsilon_i)]^{g_i}$
for the ion with bound-electrons, 
which have $M$ bound-levels $\epsilon_i$ with the degeneracy $g_i$, 
and $\lambda\!\equiv\!\exp (\beta \mu_e^0)$.
Alternatively, the grand partition function $\Xi_{\rm B}$ can 
be expanded in a polynomial of $\lambda$:
\begin{equation}\label{eq:SHgpfP}
\Xi_{\rm B} \equiv \prod_{i=1}^M[1+\lambda \exp(-\beta \epsilon_i)]^{g_i}
=\sum_{Q=0}^G\lambda^Q Z_Q
\end{equation}
with $G\! \equiv\! \sum_{i=1}^M g_i$. In this expression, 
the canonical partition function $Z_Q$ of the ion with 
the $Q$ bound electrons is defined by 
\begin{eqnarray}
Z_Q &\equiv &\sum_l \Omega(E_l^Q)\exp[-\beta E_l^Q] \label{eq:SHcpf}\\
  &=&\sum_{\sum n_s=Q}\left( \prod_{s=1}^M {g_s! 
\over n_s!(g_s-n_s)!}\right)\exp[-\beta E_{\{n_s\}}^Q]\\
&=& \sum_{\sum n_s=Q}\prod_{s=1}^M[_{g_s}C_{n_s}\exp(-\beta\epsilon_s n_s)]
\end{eqnarray}
with the total energy $E^Q_l\!\equiv\! \sum_{i=1}^M 
\epsilon_s n_s\!=\!E_{\{n_s\}}^Q$ ($n_s\!\!=$0 or 1) for the Q bound-electrons.
Here, $\Omega(E_l^Q)$ represents the number of basic states with the energy 
$E_l^Q$ for the $Q$ bound-electrons in the ion.
Furthermore, the grand partition function $\Xi_{\rm B}$ is written 
as $\Xi_{\rm B}\!=\!\sum_{i=1}^G U_Q$ using the function $U_Q$ defined by,
\begin{equation}\label{eq:SHUq}
U_Q\equiv e^{\beta \mu_{\rm e}^0 Q}Z_Q=
\sum_{\sum n_s=Q}\,\prod_{s=1}^M\{\,\,_{g_s}C_{n_s}
\exp[\,\beta (\mu_{\rm e}^0-\epsilon_s) n_s\,]\,\,\}\,.
\end{equation}
From Eqs.~(\ref{eq:SHgpfP}) and (\ref{eq:SHUq}), we obtain 
the average bound-electron number $Z_{\rm B}$ in another form:
\begin{equation}\label{eq:SHZBQ}
Z_{\rm B} = \lambda {d \over d\lambda}\ln \Xi_{\rm B}
= \sum_{Q=0}^G Q {U_Q \over \Xi_{\rm B}} =<Q>\,,
\end{equation}
which means that the probability for the bound-electron number 
of the ion to be $Q$ is given by $U_Q/\Xi_{\rm B}$.

From another point of view, let us count the number of ions 
with the $Q$ bound-electrons in a plasma, which is denoted by $N_Q$; 
this satisfies the relation $\sum_{Q=0}^G N_Q\!=\!N_{\rm I}$ 
since the total ion number in the system is $N_{\rm I}$. 
In terms of $N_Q$, the average bound-electron number $Z_{\rm B}$ 
is determined by another way:
\begin{equation}\label{eq:SHZBN}
\sum_{Q=0}^G Q {N_Q \over N_{\rm I}} =Z_{\rm B}\,.
\end{equation}
The above two expressions of Eqs.~(\ref{eq:SHZBN}) and (\ref{eq:SHZBQ}) 
for the probability that the ion in a plasma has 
$Q$ bound-electrons give rise to the relation: 
${N_Q/N_{\rm I}}\!=\!{U_Q/\Xi_{\rm B}}$, that is,
\begin{equation}\label{eq:SHansatz}
{N_Q \over U_Q} = {N_{\rm I} \over \Xi_{\rm B}}\;.
\end{equation}
Since the righ-side of this equation is independent of $Q$, we obtain 
the expression: $N_Q/U_Q\!=\!N_{Q-1}/U_{Q-1}$, 
which can be rewritten in the form:
\begin{equation}\label{eq:SHNN}
N_Q/N_{Q-1}=U_Q/U_{Q-1}=\exp[\beta \mu_{\rm e}^0] Z_Q/Z_{Q-1}\,.
\end{equation}
If we introduce the canonical partition function 
$Z'_Q \equiv \exp[\beta E_0^Q] Z_Q$ using excitation energies 
$E_n^Q\!-\!E_0^Q$ measured from the ground state $E_0^Q$ of the ion with 
the $Q$ bound-electrons, 
Eq.~(\ref{eq:SHNN}) is rewritten as 
\begin{equation}\label{eq:SH0}
{N_Q \over N_{Q-1}}=\exp[\beta(\mu_{\rm e}^0 +I_Q)]{Z'_Q \over Z'_{Q-1}}
\end{equation}
with the ionization energy $I_Q\!\equiv\! E_0^{Q\!-\!1}\!\!-\!\!E_0^{Q}$ in 
the ground state. This is the Saha equation, which is applicable to 
a plasma where the electrons may be degenerate at any degree, 
and the ions and the electrons in a plasma may interact strongly 
with each other at high densities.
When the temperature of a plasma becomes so high 
that the electrons behave as classical particles, 
the electron chemical potential is determined by the classical relation:
\begin{equation}\label{eq:SHchem}
n_0^e \lambda_e^3/2 ={n_0^e \over Z_e} = \exp[\beta \mu_{\rm e}^0]
\end{equation}
with the canonical partition function $Z_e= 2(2\pi m/h^2\beta )^{3/2}$ 
of a noninteracting electron gas and the thermal wavelength $\lambda_e$. 
As a result from the above equation, Eq.~(\ref{eq:SH0}) is reduced 
to the usual expression for 
the Saha equation\,\cite{Cowan}:
\begin{equation}\label{eq:SH1}
{n_Q  \over n_{Q-1}n_0^e} = \exp(\beta I_Q){Z'_Q \over Z'_{Q-1}Z_e}
\end{equation}
to determine the ion-density $n_Q=N_Q/V$ in the volume V.
At this point, note that the fundamental relation 
(\ref{eq:SHansatz}) to derive the Saha equation 
is nothing but an ansatz introduced by Bar-Shalom {\it et al.}\cite{Bar}.

It should be recognized here that the solution of the Saha equation, 
(\ref{eq:SH0}) or (\ref{eq:SH1}), is 
obtained by determining the partition function $U_Q$ of a plasma 
with aids of the relation (\ref{eq:SHansatz}), 
that is $n_Q=n_0^{\rm I} U_{Q}/\Xi_{\rm B}$.
For this calculation, we can use the following recursive formulas\cite{Bar,Ble} 
for the partition function $U_Q$ of the ion with the $Q$ bound-electrons:
\begin{eqnarray}
&&U_0=1 \label{eq:SHrc1}\;,\\
&&U_Q=\sum_{n=1}^Q\chi_n U_{Q-n}/Q \label{eq:SHrc2}\;,
\end{eqnarray}
where $\chi_n\!=\!-\sum_{i=1}^M g_i(-X_i)^n$ 
with $X_i\!=\!\exp[-\beta (\epsilon_i\!-\!\mu_e^0)]$.

An applied example of our formula to evaluate the ion population $P(Q)$
is shown in the case of Rb plasmas.
In the nucleus-electron model based on the DF theory, the bound-levels $\epsilon_i$ of 
the ion in a plasma is determined by solving the wave-equation 
for the self-consistent potential given by $U_{\rm eI}^{\rm eff}(r)$ (\ref{eq:Uei}), 
and the chemical potential $\mu_e^0$ of electrons is evaluated 
by the condition (\ref{eq:chem}); the temperature variation of 
the bound-levels of Rb plasma was shown in Fig.~\ref{fig:levels} 
for a range from 0 eV to 30 eV.
Using these values, we can obtain $U_Q$ from the 
recursive relations, (\ref{eq:SHrc1}) and (\ref{eq:SHrc2}). In this way, 
the charge population $P(Q)\equiv U_Q / \Xi_{\rm B}$ is evaluated 
for Rb plasma with the atomic number $Z_{\rm A}\!\!=\!\!37$ 
varying the temperature from 3 eV to 30 eV at fixed ion-density 
$1.03\!\times\! 10^{22}/{\rm cm}^3$ ($r_{\rm s}^{\rm I}=5.388$), 
as was studies in \S.~\ref{num} for the evaluation of the 
electron-ion and ion-ion RDF's. Figure~\ref{fig:pop} 
displays the charge population $P(Q)$ in 
Rb plasmas for this temperature variation; the bound-electron 
number $Z_{\rm B}$ from Eq.(\ref{eq:SHZBN}) 
is coincident with the values obtained previously by Eq.~(\ref{eq:shzb}) 
for each temperature, as a matter of course. At a sufficiently 
low temperature such as 
1 eV, the charge population reduces to $P(Q)=\delta_{Q,36}$ 
for Rb plasmas at liquid-metal density.

\section{CONCLUSIVE DISCUSSION}
\begin{figure}
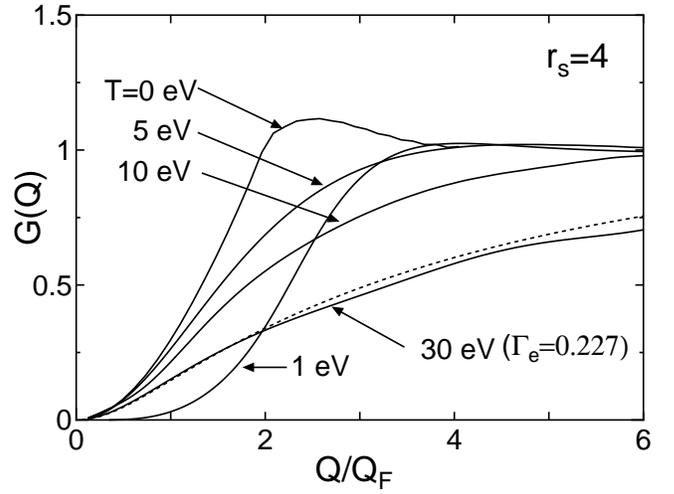

\caption{The local-field corrections (LFC) at temperatures 
from 0 to 30 eV determined by the one-component QHNC equation.
The LFC calculated by the classical HNC for the OCP with $\Gamma_e\!=\!0.227$ is 
displayed by the dashed curve, which indicates that 
this can be used to approximate the QHNC LFC at this temperature.}
\label{fig:Gq}
\end{figure}

\begin{figure}
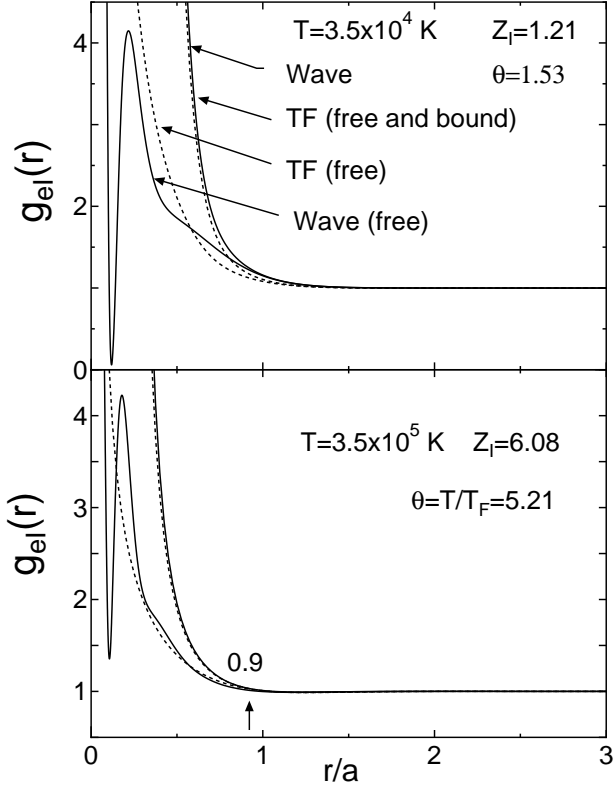

\caption{The electron-ion RDF's calculated from the wave equation 
and the TF approximation at temperatures of $3.5\times 10^4$ K (3 eV) and 
$3.5\times 10^5$ K (30 eV).
The solid curves denote the results from the wave equation 
and the dashed curves, from the TF approximation. 
The TF approximation can not give a good description of the 
free-electron density distribution $n_e^{\rm f}(r|\rm I)\!=\!n_0^eg_{\rm eI}(r)$ 
in the core-region even at a high temperature (30 eV), 
where the total (bound and free) electron-density distribution is 
fairly well described by the TF approximation. 
The electron-ion RDF can be approximated by the TF formula for 
larger distances than the point (0.9) denoted by the arrow.}
\label{fig:TFg}
\end{figure}

We have demonstrated that the QHNC method, which has been successfully applied 
to many kinds of liquid metals, can be extended to treat a partially ionized plasma, 
taking rubidium as an illustrating example. In the application of 
the QHNC method to a plasma, it is necessary to use the LFC $G(Q)$ 
at arbitrary temperature, which is determined by the one-component 
QHNC equation for an electron gas in the jellium; the QHNC $G(Q)$ 
reduces to the LFC obtained by the classical HNC for the OCP 
at high temperatures as is shown in Fig.~\ref{fig:Gq}. 
In the numerical calculation in the QHNC method for a plasma, 
it is very time-consuming to evaluate free-electron density 
distributions at high temperatures; this problem can be easily 
circumvented by combined use of the TF approximation as discussed 
in \S.~\ref{num}. However, the simple TF approximation to 
the free-electron density distribution is shown to give 
only a rough estimation in Fig.~\ref{fig:TFg}; 
this may not be applied to calculate the accurate atomic structure 
in a plasma, although there are many examples as mentioned before.

\begin{figure}
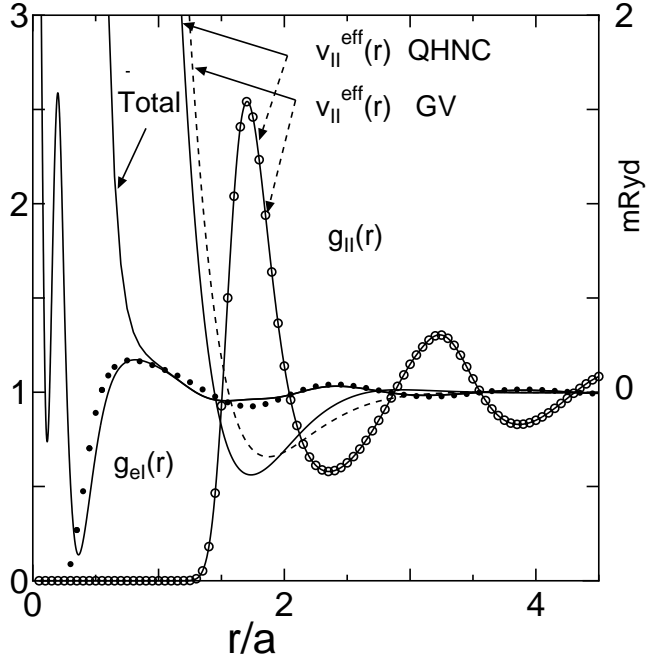

\caption{The electron-ion and ion-ion RDF's with the effective 
interactions in liquid Rb at 313 K. 
The solid curves denote the RDF's, the total electron-density 
distribution and the effective ion-ion interaction 
calculated by using the QHNC $G(Q)$. 
The dashed curve designates the effective ion-ion potential 
based on the Geldart-Vosko $G(Q)$, which yields the ion-ion RDF 
plotted by the open circles; $\bullet$,
the electron-ion RDF derived by using an Ashcroft potential. }
\label{fig:iieiRDF}
\end{figure}

\begin{figure}
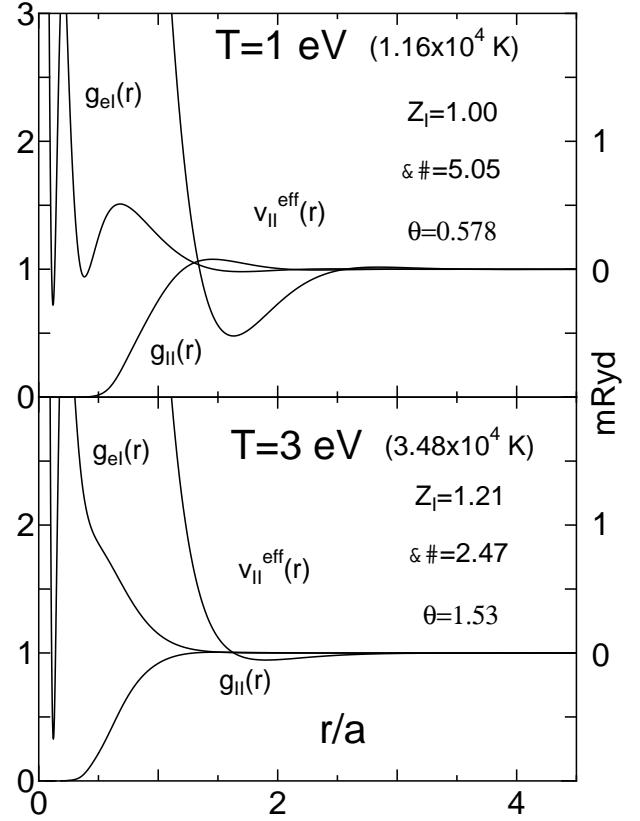

\caption{The electron-ion and ion-ion RDF's together with
the effective ion-ion interaction at temperatures of 1eV and 3 eV. 
The electron-ion RDF at 1 eV has an inner-core structure similar to the 
RDF of a liquid state at the normal condition (cf. Fig.~\ref{fig:iieiRDF}), 
and this inner-core structure near $r/a\!=\!0.4$ disappears 
at 3 eV with a significant ionization $Z_{\rm I}\!=\!1.21.$
}
\label{fig:1ev3ev}
\end{figure}

In a liquid metal, the electron-ion RDF has a clear 
inner-core structure distinct from outer-core structure 
(cf. Fig.~\ref{fig:iieiRDF}); this distinction enables us 
to construct a pseudopotential in a liquid liquid metal. 
In the case of Rb, it remains as a liquid metal even 
at a temperature of $1.16\!\times\!10^4$ K where the inner- and 
outer-core distinction is clearly seen with $Z_{\rm I}\!=\!1$ as 
is displayed in Fig.~\ref{fig:1ev3ev}; 
this distinction disappears at 3 eV, where rubidium becomes 
a plasma with a significant ionization $Z_{\rm I}\!=\!1.21$. 

It is important to remember that the ion-sphere (IS) model   
is not an appropriate approximation to treat a high-temperature plasma with 
a small plasma-parameter $\Gamma$.
In the IS model, the ion-ion RDF is approximated by the 
step function with the Wigner-Seitz radius $a$. Therefore, this approximation 
is only valid in the strongly correlated region, 
where the structure factor $S_{\rm II}(0)\!\approx \!0$ at zero wavenumber, because of the relation:
\begin{equation}\label{eq:ISM}
-n_0^{\rm I}\int[g_{\rm II}(r)-1]d{\bf r}
=1 - S_{\rm II}(0)\approx n_0^{\rm I}4\pi a^3/3\equiv 1\;.
\end{equation}
This condition is very important to keep charge neutrality 
around the ion\cite{chi90}:
\begin{equation}\label{eq:Cneut}
Z_{\rm I}=-Z_{\rm I}n_0^{\rm I}\int[g_{\rm II}(r)-1]d{\bf r}
 + n_0^{\rm e}\int[g_{\rm eI}(r)-1]d{\bf r}\,.
\end{equation}
When the ion-ion RDF becomes weaker as the temperature increases, 
the structure factor $S_{\rm II}(0)$ grows large; there, 
the IS model is not properly applicable. 
This situation is exemplified in our calculation of 
the ionization $Z_{\rm I}$. 
The IS (jellium-vacancy) model provides the ionization $Z_{\rm I}$: 
1.27, 2.05, 3.80, 5.28 and 6.55 for temperatures: 3, 5, 10, 22, 
and 30 eV, respectively, while the corresponding values from 
the fully self-consistent QHNC calculation are 
1.21, 1.96, 3.71, 5.08 and 6.08, respectively. 
We can see that significant differences are manifested between the 
IS and QHNC results, as the temperature increases. The reason for 
these differences can be ascribed to the fact that the IS model produces 
a screening effect from the ions contained in Eq.~(\ref{eq:Cneut}) 
in the approximated form $Z_{\rm I}[1\!-\!S_{\rm II}(0)]\!\approx\!Z_{\rm I}$, 
which becomes too strong for a plasma at high temperatures 
where $S_{\rm II}(0)$ becomes large.
As a consequence, the neutral-atom model\cite{NPA} based on 
the IS model is, also, not appropriate to construct 
an effective ion-ion interaction at a high-temperature plasma.
As discussed in \S.\ref{num}, the IS model can be improved by using the 
ion-ion RDF from the HNC equation for a screened Debye potential instead of 
the step function, in dealing with a high-temperature plasma.

\begin{figure}
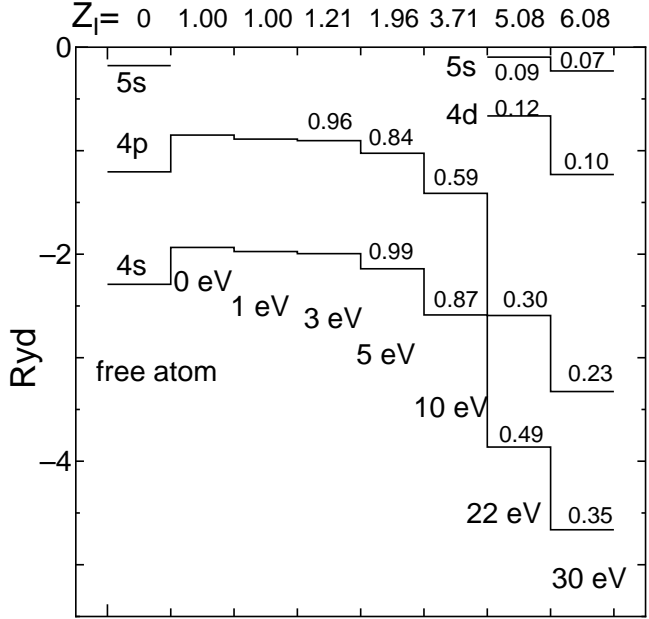

\caption{The temperature variation of outer bound levels (4s, 4p, 4d and 5s) 
in the Rb ion in a plasma for a range from 0 eV to 30 eV at a  
fixed density of $r_{\rm s}^{\rm I}=5.388$. 
The bound levels become shallow when free atoms are compressed 
to be a liquid state (0 eV), and turn to become deeper as the temperature 
is increased with the fixed ion-density.
Numbers attached to bound levels
denote the occupation numbers $f(\epsilon_i)$, and the ionization $Z_{\rm I}$ 
for each temperature is written in the top of this figure.
}
\label{fig:levels}
\end{figure}

\begin{figure}
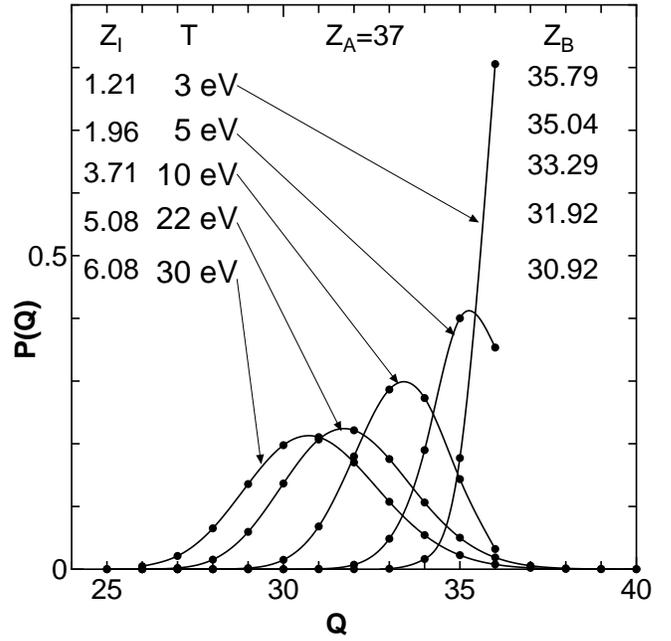

\caption{The dependence of the charge population $P(Q)$ on 
the temperature varying from 3 eV to 30 eV for Rb plasma 
($Z_{\rm A}\!\!=\!\!37$) at a fixed ion-density of 
$1.03\!\times\!10^{22}/{\rm cm^3}\,\,(r_{\rm s}^{\rm I}\!=\!5.388)$.
This charge population provides the average bound-electron 
$Z_{\rm B}$ of ion, as denoted in this figure for each 
temperature along with the ionization $Z_{\rm I}\!\equiv\! Z_{\rm A}\!-\!Z_{\rm B}$.
}
\label{fig:pop}
\end{figure}

In the QHNC method, the inner-structure (the atomic structure 
of the ion) is determined in the consistent way with the outer 
structures (the electron-ion and ion-ion RDF's and the average 
ionization $Z_{\rm I}$). Therefore, we can expect this method to 
provide an accurate procedure to deal with the atomic structure 
in a high-density plasma; the bound-levels in an ion can be calculated 
by taking account of the density and temperature effects as is 
shown in Fig.~\ref{fig:levels}. In addition, it should be remarked 
that the DF theory leads to the Saha equation as discussed 
in \S.~\ref{saha}, and the QHNC method based on the DF theory can 
provide a procedure to solve the Saha equation with ease 
by using the recursive formula. As an applied example of this formula, 
the charge population $P(Q)$ is calculated 
from the QHNC result for a Rb plasma, as is displayed in Fig.~\ref{fig:pop}.
Moreover, we can expect that the QHNC method can  solve various 
kinds of problems associated with the atomic structure in a plasma 
by taking account of the plasma effects. For example, with the combined 
use of Slater's transition-state method\cite{TS}, we have already 
calculated the shift-variation of the K-edge\cite{KedgeUnpub} in an aluminum plasma 
along the shock Hugoniot in good agreement with the experiment 
performed by DaSilva {\it et al.}\cite{DaS}. 

The QHNC method can provide an accurate description of 
the metallic system for a wide range of densities and temperatures 
from the liquid-metal to the plasma state in a unified manner, 
as is ascertained from many experiments on liquid metals. 
This fact indicates that the QHNC method can be used to calculate 
transport properties and equation of state in a wide region 
from the liquid-metallic to the plasma state, where there has been 
no systematic applicable theory up to the present.

With decreasing temperature or increasing pressure of a plasma, 
some bound state of each ion in a liquid metal or plasma begins 
to disappear into the continuum; it becomes a narrow-resonant state and 
disappears gradually as a wide resonance in the continuum.
In our calculation of plasma states we do not take account of 
resonant states. 
In practice, the resonant-state contribution in a plasma is not as significant as 
in the case of a liquid metal such as a transition metal.
A precise definition of a resonant state \cite{n60} is given by the 
pole $\widetilde 
E_{n\ell}$ of 
the S-matrix $S_{\ell}(E)$ concerning the wave equation for an 
electron under the effective potential (\ref{eq:df-pot}) with 
$S_{\ell}(E)=\exp(2i\delta_{\ell}(E))$ for phase shifts $\delta_{\ell}(E)$.   
In a strict way, it is required \cite{QHNC} that the 
\lq\lq bound"-electron number $Z_{\rm B}$ in an ion should include a 
contribution of the physical resonant states ($\:|\Im m \widetilde 
E_{n\ell}|\ll\Re e \widetilde E_{n\ell}\:$) in addition to the bound 
electrons with $\epsilon_i<0$ in such a way that
\begin{equation}
Z_{\rm B}\equiv \sum_{\epsilon_i<0}f(\epsilon_i)+\sum_{n\ell\in {\rm 
phys.res.}}2(2\ell+1)\Re e[F(\widetilde E_{n\ell})] \;. \label{e:gzb} 
\end{equation}
In the above, $F(\widetilde E_{n\ell})$ is the function introduced 
by More\cite{m85} with the definition: 
\begin{equation}
F(\widetilde E_{n\ell})\equiv
\frac1{i\pi}\int_0^{\infty}\left({E \over \widetilde
E_{n\ell}}\right)^{1/2}\!\!\!\frac{f(E)}{E-\widetilde E_{n\ell}}dE\;, 
\label{e:gfd}
\end{equation}
to represent the thermal occupation probability of 
a resonant state $\widetilde E_{n\ell}$.
Also, the chemical potential $\mu_{\rm e}^0$ in consideration of 
the resonant states should be determined by 
\begin{eqnarray}
Z_{\rm A}=&&\sum_{\epsilon_i<0} {1\over \exp[\beta 
(\epsilon_i-\mu_{\rm e}^0)]+1} +\sum_{n\ell\in {\rm phys.res.}}2(2\ell+1)\Re
 e[F(\widetilde E_{n\ell})] \nonumber \\&&+\frac1{n_0^{\rm I}}\int {2\over 
\exp[\beta (p^2/2m-\mu_{\rm e}^0)]+1} 
\frac{d{\bf p}}{(2\pi\hbar)^3}\;. \label{e:gcp}
\end{eqnarray}
However, the determination of the charge occupation $P(Q)$ 
taking account of the resonant-state contribution is a problem which 
remains to be investigated. 

Molecular-dynamics (MD) simulation is necessary to study complex 
systems which are 
inhomogeneous or time dependent and so on. In the MD simulation of a dense
plasma, the Coulomb interactions among close and
distant particles must be calculated precisely and efficiently; 
the Particle-Particle Particle-Mesh (PPPM) method\cite{pic,Nishi} 
should be used in the simulation code to treat many particles. In the 
particle-particle method the Coulomb forces among close particles are 
directly summed up and in the particle-mesh method the forces on a particle 
are interpolated from electric fields at mesh points. In the SCOPE 
(Strongly COupled Plasma ParticlE) code\cite{Nishi,Furu,Ueshi} based on the PPPM method, 
the Deutsch potential
\cite{deut} is adopted to imitate quantum effects. With use of the code, 
bremsstrahlung emission\cite{Furu},
transport coefficients and the Lyapunov exponents\cite{Ueshi} were obtained in
dense plasmas.
However, the applicability of the Deutsch potential is limited to 
a hydrogen plasma or to a fully ionized plasma at most, since the ion structure is 
not considered in the derivation of the Deutsch potential. 
In order to perform a classical MD simulation (SCOPE) on a partially ionized plasma, 
we must introduce effective classical potentials 
$v_{ij}^c(r)$ applicable to partially ionized ions 
in a plasma as a classical mixture of ions and electrons; 
the QHNC method can produce these effective potentials as follows.
The quantum effects of electron-electron interaction can be taken 
into account by defining 
an effective classical pair-potential $v_{\rm ee}^c(r)$ between electrons 
in such a way that the HNC equation 
for $n_{e}^c(r|e)\!=\!n_0^eg_{\rm ee}^c(r)$ in classical fluids 
with $v_{\rm ee}^c(r)$ provides the same electron-density 
distribution $n_{e}^{\rm QHNC}(r|e)$ determined by the one-component 
QHNC equation (\ref{eq:QHNCee}); 
this condition is written in the following integral equation for $v_{\rm ee}^c(r)$,
\begin{equation}\label{eq:vee-c}
n_{e}^c(r|e)\equiv n_0^e\exp[-\beta v_{\rm ee}^c(r)\!+
\!\gamma^c(r)]=n_{e}^{\rm QHNC}(r|e)
\end{equation}
with
$\gamma^c(r)\equiv \int C_{\rm ee}^c(|{\bf r}\!-\!{\bf r}'|)
[n_{e}^c(r'|e)\!-\!n_0^e]d{\bf r}'\,.$ 
 In a similar way, an electron-ion classical potential 
$v_{\rm eI}^c(r)$ is determined by the condition that the classical 
electron-ion RDF $g_{\rm eI}^c(r)$ should be  identical with the QHNC result:
\begin{equation}\label{eq:vei-c}
g_{\rm eI}^c(r)\equiv 
\exp[-\beta v_{\rm eI}^c(r) +{\it \Gamma}_{\rm eI}^c(r)]
= g_{\rm eI}^{\rm QHNC}(r) 
\end{equation}
with
\begin{equation}\label{eq:Gamme-eIc}
{\it\Gamma}_{\rm eI}^c(r)\equiv 
\int C_{\rm ee}^c(|{\bf r}\!-\!{\bf r}'|)n_0^{\rm e}
[g_{\rm eI}^c(r')\!-\!1]d{\bf r}'
+\int C_{\rm eI}^c(|{\bf r}\!-\!{\bf r}'|)n_0^{\rm I}
[g_{\rm II}^{\rm QHNC}(r')\!-\!1]d{\bf r}\;.
\end{equation}
With use of the effective potentials determined above, the SCOPE code 
can be applied to investigate dynamical problems in a partially 
ionized plasma as a classical ion-electron mixture.

We have shown that the QHNC method is extended to treat 
a {\it partially ionized} plasma in a wide range of densities and 
temperatures, and provides the average ionization, 
the electron-ion and ion-ion RDF's, the atomic structure of 
the ions and the charge population of differently ionized species 
in a self-consistent manner from the atomic number as the only input data.
Therefore, this method produces the fundamental quantities 
necessary to calculate the plasma properties, 
and offers a procedure to treat the spectroscopic problem in a plasma. 
It should be kept in mind that the QHNC method can provide 
a precise description of \lq\lq simple" plasma where the bound states 
are clearly distinguished from the continuum state; to take 
into account the resonant states in a plasma, 
some improvement is necessary as was discussed in the previous work\cite{QHNC}.
\acknowledgments
J.C. would like to thank Dr. F. Perrot for providing 
his accurate subroutines necessary for our extension of 
the QHNC code to treat a plasma.

%
%

%
%

\end{document}